\documentclass[aps,prx,showpacs,preprintnumbers,twocolumn,groupedaddress]{revtex4}
\usepackage{graphicx}
\usepackage{color}

\begin{document}

\title{Quantum Monte Carlo simulation of three-dimensional Bose-Fermi mixtures}
\preprint{RIKEN-QHP-46}
\author{Arata~Yamamoto$^1$}
\author{Tetsuo~Hatsuda$^{1,2}$}
\affiliation{$^1$Theoretical Research Division, Nishina Center, RIKEN, Saitama 351-0198, Japan\\
$^2$IPMU, The University of Tokyo, Kashiwa 277-8583, Japan}

\date{\today}

\begin{abstract}
 Exploratory simulations of Bose-Fermi mixtures on the three-dimensional 
 optical lattice at finite temperature are performed by adopting the 
 lattice quantum chromodynamics technique.
 We analyze the bosonic superfluid transition 
 and its dependence on the strength of the boson-fermion coupling.
  The particle densities and  the pair occupancies are also
  studied to understand the effect of the boson-fermion coupling to the 
  microscopic properties of the system. Effect of the induced  fermion-fermion interaction
   by the boson density fluctuation is clearly seen.
\end{abstract}


\maketitle

\section{Introduction}
Many-body mixture of bosons and fermions is an
interesting quantum system.  It is not only useful to study the 
interplay between  different quantum statistics in condensed matter physics
and in atomic physics \cite{Pethick:2008},
 but also important for understanding the hadron-quark
 phase transition in nuclear and particle physics \cite{Maeda:2009ev}.
% The most famous example of Bose-Fermi mixtures is a $^4$He-$^3$He mixture.
Recently, many kinds of Bose-Fermi mixtures have been realized
 in ultracold atomic experiments. 
In particular,the  Bose-Fermi mixtures 
 such as  $^{87}$Rb-$^{40}$K \cite{Gunter:2006,Ospelkaus:2006,Best:2009zz}, $^{170}$Yb-$^{173}$Yb and $^{174}$Yb-$^{173}$Yb \cite{Sugawa:2011}
  can be trapped in three-dimensional optical lattices. 
  
 Despite of its importance,   ab initio quantum Monte Carlo 
 calculations of the Bose-Fermi Hubbard model, which is a model system of the  
 Bose-Fermi mixture on the optical lattice,    have been limited to  one spatial
  dimension \cite{Takeuchi:2006,Pollet:2006,Sengupta:2007,Hebert:2007,Pollet:2008,Varney:2008,Masaki:2008}. 
(See also a recent review \cite{Pollet:2012}.)
 These simulations were done with improved versions of the world-line formalism \cite{Hirsch:1983wk,Prokofev:1998,Sandvik:1999}.
 The world-line formalism is an exact scheme in one spatial dimension while it has the fermion sign problem in higher dimensions.
%The exchange of fermion positions leads to a minus sign, and results in the sign problem.

In this paper, 
 we apply the method of lattice Monte Carlo simulation of quantum chromodynamics (QCD)
  \cite{Rothe:2005nw} to overcome the fermion sign problem 
  in the Bose-Fermi mixture of ultracold atoms
 on an optical lattice in  three spatial dimensions at finite temperature.
 The lattice QCD simulation is an established numerical framework for the strongly coupled 
 regime of QCD; the mixture of gluons (bosons) and quarks (fermions) can be treated in 
 any dimension 
 without fermion sign problem as long as the fermion determinant is real and semi-positive.
 (For relativistic fermions with finite chemical potential, 
 the fermion determinant becomes complex, while the chemical potential for 
  non-relativistic fermions does not cause the problem.)
%the fermion sign problem is exactly avoided.
%with a finite quark chemical potential, the fermion determinant is complex. 
% The simulation suffers from the fermion sign problem.)}

 This paper is organized as follows.
In Sec.~II, we introduce the model action and discuss the basic setup of our simulations.
 In Sec.~III, we analyze the bosonic superfluid  transition
 and its dependence on the strength of the boson-fermion coupling.
  The particle densities and  the pair occupancies are also
  studied to understand the effect of the boson-fermion coupling.
 In Sec.~IV, we summarize this study.  
A part of the preliminary results was reported in Ref.~\cite{Yamamoto:2012bj}.

\section{Formulation}
We consider one-component boson field $\Phi(\vec{x},\tau)$ and two-component fermion field $\Psi_\uparrow(\vec{x},\tau)$ and $\Psi_\downarrow(\vec{x},\tau)$.
The partition function $Z$ is then written as a path integral in terms of  the
 Euclidean action $S$:
\begin{eqnarray}
Z &=& \int D\Phi^* D\Phi D\Psi_\uparrow^* D\Psi_\uparrow D\Psi_\downarrow^* D\Psi_\downarrow \ e^{-S} \nonumber \\
&=&  \int D\Phi^* D\Phi \ \det K_\uparrow \det K_\downarrow e^{-S_B} \ .
\label{eq:Z}
\end{eqnarray}
In the second line, the bilinear form of the
fermion fields are integrated out, so that
 $Z$ becomes a functional only of the boson field.

We adopt the single-band Bose-Fermi Hubbard model whose 
 lattice action reads \cite{Yamamoto:2012bj,Hasenfratz:1983ba,Chen:2003vy}
\begin{eqnarray}
S = S_B +S_F +S_{BF},
\end{eqnarray}
\begin{widetext}
\begin{eqnarray}
\label{eq:SB}
S_B &=& \sum_{\vec{x},\tau} \Big[ \Phi^*(\vec{x},\tau) \{ \Phi(\vec{x},\tau) - e^{\mu_B} \Phi(\vec{x},\tau-1) \}  -  \sum_{j=1}^3 t_B \{ \Phi^*(\vec{x},\tau) \Phi(\vec{x}+\vec{e}_j,\tau)+ \Phi^*(\vec{x},\tau) \Phi(\vec{x}-\vec{e}_j,\tau) \} \nonumber \\
&&+ U_{BB} \Phi^*(\vec{x},\tau)\Phi(\vec{x},\tau) \{\Phi^*(\vec{x},\tau)\Phi(\vec{x},\tau) -1\} \Big] \\
\label{eq:SF}
S_F &=& \sum_{\vec{x},\tau,\sigma} \Big[ \Psi_\sigma^*(\vec{x},\tau) \{ \Psi_\sigma(\vec{x},\tau) - e^{\mu_F} \Psi_\sigma(\vec{x},\tau-1) \} -  \sum_{j=1}^3 t_F \{ \Psi_\sigma^*(\vec{x},\tau) \Psi_\sigma(\vec{x}+\vec{e}_j,\tau) + \Psi_\sigma^*(\vec{x},\tau) \Psi_\sigma(\vec{x}-\vec{e}_j,\tau) \} \Big] \\
\label{eq:SBF}
S_{BF} &=& \sum_{\vec{x},\tau,\sigma} U_{BF} \Phi^*(\vec{x},\tau)\Phi(\vec{x},\tau) \Psi_\sigma^*(\vec{x},\tau)\Psi_\sigma(\vec{x},\tau-1).
\end{eqnarray}
\end{widetext}
%The two pseudospin components of the fermion field are degenerated.
Here, $\mu_B$ ($\mu_F$) is the boson (fermion) chemical potential, 
 $t_B$ ($t_F$) is the 
 spatial hopping parameters for the boson (fermion). Also, 
 $U_{BB}$ and $U_{BF}$ are the strength of boson-boson interaction and 
  boson-fermion interaction, respectively.
 We do not consider the fermion-fermion interaction for simplicity, i.e.  $U_{FF}=0$.
The temporal hopping structures of the chemical potential and the interaction are based on Refs.~\cite{Hasenfratz:1983ba,Chen:2003vy}.

 Note that we have absorbed the spatial lattice spacing $a_s$ and the
  temporal lattice spacing $a_\tau$ into the definition of the fields and
  the coupling parameters.  Here $a_s$  should be identified with a  
 finite lattice constant of an optical lattice, while $a_\tau$ should be
 eventually taken to be zero to recover the continuum
  limit in the temporal direction. 
In the following discussion, all dimensional quantities are scaled by the lattice constant.

  Our lattice action has the boson sign problem which appears in non-relativistic systems. 
 The boson action is complex in general because the temporal hopping term is given as 
   $\sum_{\tau} \Phi^*(\vec{x},\tau) \{ \Phi(\vec{x},\tau) - \Phi(\vec{x},\tau-1) \} 
   = N_\tau \sum_{\nu} \{ 1-e^{-i\omega_B(\nu)} \} 
   \tilde{\Phi}^*(\vec{x},\nu) \tilde{\Phi}(\vec{x},\nu)$ 
   with the boson Matsubara frequency $\omega_B(\nu) = 2 \nu \pi T$.
To avoid this boson sign problem, we adopted the zero-frequency approximation,
in which the boson field is projected onto the zero-frequency mode as $\Phi(\vec{x},\tau) = \tilde{\Phi}(\vec{x},\nu=0) \equiv \Phi(\vec{x})$.
This approximation is validated if the system is either in the 
  critical region or  in the high-temperature limit, where
  the static mode of the  boson field, $\nu=0$,
  gives a dominant contribution.
Furthermore, effect of the non-zero Matsubara modes 
can be estimated by using the reweighting method \cite{Ferrenberg:1988yz}.

We can prove that the fermion determinant is real and positive.
Let us consider the fermion action in the frequency space.
The matrix $K_\sigma$ becomes diagonal in the frequency space.
The complex factor is only $e^{-i\omega_{F}(\nu)}$ with the fermion Matsubara frequency $\omega_F(\nu) = (2 \nu-1) \pi T$.
The fermion determinant is
\begin{eqnarray}
\det K_\sigma &=& \prod_\nu (A+ e^{-i\omega_{F}(\nu)}B) \nonumber \\
&=& \prod_{\omega_{F}(\nu)>0} [ \{A+ B\cos \omega_{F}(\nu)\}^2 + B^2 \sin^2 \omega_{F}(\nu) ] \nonumber \\
&>& 0 \ ,
\end{eqnarray}
where $A$ and $B$ are real numbers.
Therefore, there arises
 no sign problem originating from the fermion determinant.
 (This is in contrast to the Bose-Fermi system with a Yukawa coupling 
 discussed e.g. in Ref.~\cite{Scalapino:1981}.)
Even if the fermion-fermion interaction is introduced, there is no fermion sign problem:  After the Hubbard-Stratonovich transformation of the fermion-fermion interaction,
  $K_\sigma$ is still a real matrix  with auxiliary fields, so that 
 the two-component fermion determinant becomes semi-positive, 
$ \det K_\uparrow \det K_\downarrow = (\det K_\uparrow)^2 \ge 0$.

 Our numerical simulation procedure is based on 
  the hybrid Monte Carlo algorithm with the pseudofermions \cite{Duane:1987de},
  which is widely used in modern simulations of lattice field theories.
   For preconditioning the fermion matrix, we perform a Fourier transformation from the imaginary time $\tau$ to the fermion Matsubara frequency $\omega_F(\nu)$.
In this study, we fixed the hopping parameters $t_B = t_F = 0.01$ and the repulsive boson-boson
 interaction $U_{BB} = 0.1$ for the first trial.
We performed the simulation in a spatial lattice volume $V=N_s^3 = 10^3$ with periodic boundary conditions.
We varied temperature $T=1/N_\tau$ by changing the temporal lattice size  as
$N_\tau= 20$, 30, 40, 60, and 100.

In the grand-canonical approach,
 particle number densities are related to chemical potentials and other parameters.
 Therefore, 
 we monitor the boson number density $n_B$ and fermion number density $n_F$ 
 in our simulations: 
\begin{eqnarray}
n_B &=& \langle \Phi^*(\vec{x}) \Phi(\vec{x}) \rangle , \\
n_F &=& \langle \Psi_\uparrow^*(\vec{x},\tau) \Psi_\uparrow(\vec{x},\tau) \rangle = \langle \Psi_\downarrow^*(\vec{x},\tau) \Psi_\downarrow(\vec{x},\tau) \rangle .
\end{eqnarray}
Note that, in the mean-field level, the
 boson-fermion coupling $U_{BF}$ induces effective chemical potentials:  
\begin{eqnarray} 
 \mu_B^{\rm eff} &\sim & \mu_B - U_{BF}n_F, \label{eq:muBeff}\\
 \mu_F^{\rm eff} &\sim &  \mu_F - U_{BF}n_B.\label{eq:muFeff}
\end{eqnarray}
 This implies that attractive boson-fermion interaction ($U_{BF} < 0$)
 increases the effective chemical potentials and hence the particle number densities.
 Our numerical simulations given below indeed  show such behavior.

\section{Numerical results}

\begin{figure}[b]
\begin{center}
\includegraphics[scale=1]{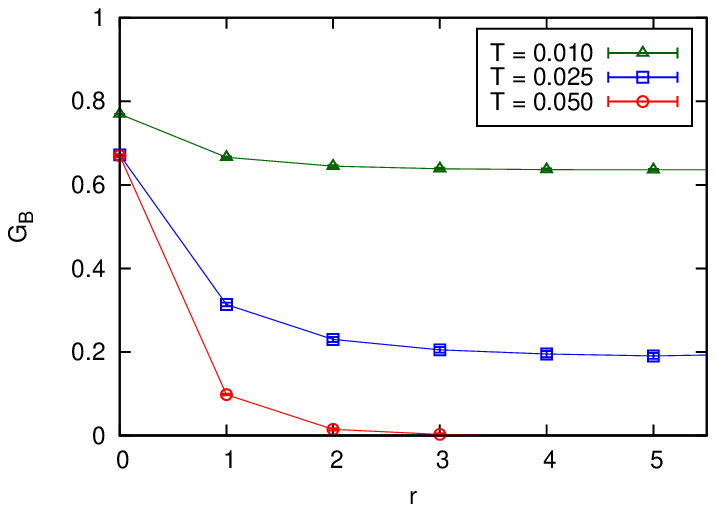}
\caption{\label{fig1}
The boson propagator $G_B(r)$ as a function of 
 $r= |\vec{x}-\vec{y}|$ for different values of temperature $T$ with $U_{BF}=0$  and $\mu_F=\mu_B=0$.
 }
\end{center}
%\end{figure}

%\begin{figure}[b]
\begin{center}
\includegraphics[scale=1]{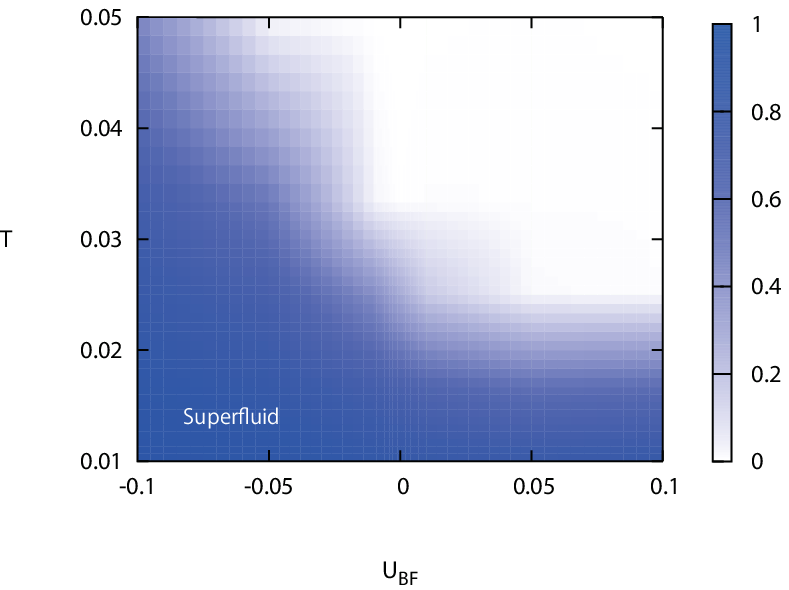}
\caption{\label{fig2}
The bosonic condensate fraction $R_B$ in the $T$-$U_{BF}$ plane at $\mu_F=\mu_B=0$.
 }
\end{center}
\end{figure}

To see the effect of $U_{BF}$ on the Bose-Einstein condensation,
let us first examine  the boson propagator
 as a function of the distance $r=|\vec{x}-\vec{y}|$,
\begin{eqnarray}
 G_B(r)= \langle \Phi^* (\vec{x}) \Phi (\vec{y}) \rangle \ .
\end{eqnarray} 
The boson propagator $G_B(r)$ with $U_{BF}=0$ and $\mu_F=\mu_B=0$ is shown in  Fig.~\ref{fig1}.
Although the statistical errors are shown by the error-bars in the figure,  they are smaller than the size of the symbols and practically invisible.
 At high temperature ($T >  0.02$), $G_B(r)$  drops to zero exponentially,
 while at low temperature ($ T < 0.02$),
 $G_B(r)$ approaches to a constant value for large $r$, i.e. a signature of the
the off-diagonal long-range order.

In Fig.~\ref{fig2}, we draw the boson condensate fraction defined by
\begin{eqnarray}
R_B= 
\frac{G_B(r=N_s/2)}{G_B(r=0)} \ , 
\end{eqnarray}
for different values of $T$ and $U_{BF}$ at $\mu_F=\mu_B=0$.
The blue region corresponds to the superfluid phase.
The figure shows that  
  the superfluid region is enlarged  as the boson-fermion attraction ($U_{BF}<0$)
    becomes strong.
  This is because  $\mu_B^{\rm eff}$ and hence the boson number density $n_B$
    increase associated with the increase of the  
     boson-fermion attraction. 

\begin{figure}[t]
\begin{center}
\includegraphics[scale=1]{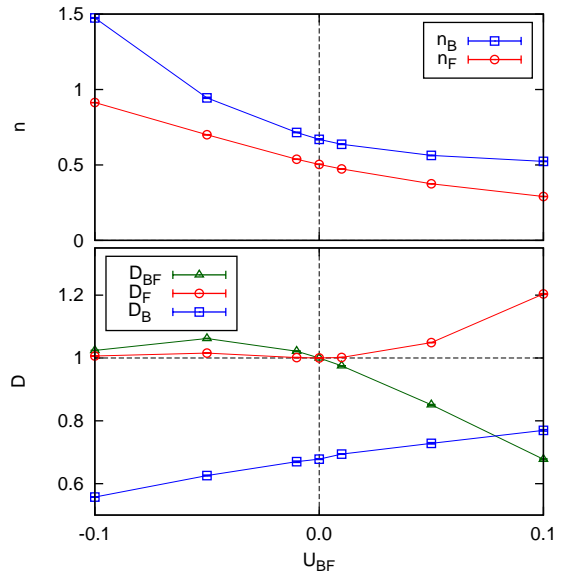}
\caption{\label{fig3}
The particle number densities and pair occupancies  at $\mu_B=\mu_F=0$ and $T=0.05$ as a function of $U_{BF}$. 
}
\end{center}
\end{figure}

To see the effect of $U_{BF}$ on particle number profiles in more details,
let us introduce normalized pair occupancies 
 which is a measure to find a pair at the same lattice site;
\begin{eqnarray}
D_{BF} &=& \frac{\langle n_B (\vec{x}) n_{F\sigma} (\vec{x}) \rangle}{\langle n_B (\vec{x}) \rangle \langle n_{F\sigma} (\vec{x}) \rangle} \label{eq:DBF}\\
D_F &=& \frac{\langle n_{F\uparrow} (\vec{x}) n_{F\downarrow} (\vec{x}) \rangle}
{\langle n_{F\uparrow} (\vec{x}) \rangle \langle n_{F\downarrow} (\vec{x}) \rangle} \label{eq:DFF}\\
D_B &=& \frac{1}{2} \frac{\langle n_B (\vec{x}) n_B (\vec{x}) \rangle}
{\langle n_B (\vec{x}) \rangle^2} \ . \label{eq:DBB}
\end{eqnarray}
Here the number density operators are  defined as 
$n_B (\vec{x}) \equiv \Phi^*(\vec{x}) \Phi(\vec{x})$ 
and $n_{F\sigma} (\vec{x}) \equiv \Psi_\sigma^*(\vec{x},\tau) \Psi_\sigma(\vec{x},\tau)$.
The coefficient 1/2 in $D_B$ is a symmetry factor for identical particles.
 The pair occupancies are experimentally observed
 on a three-dimensional optical lattice \cite{Sugawa:2011}.
 
   \begin{figure}[t]
\begin{center}
\includegraphics[scale=1]{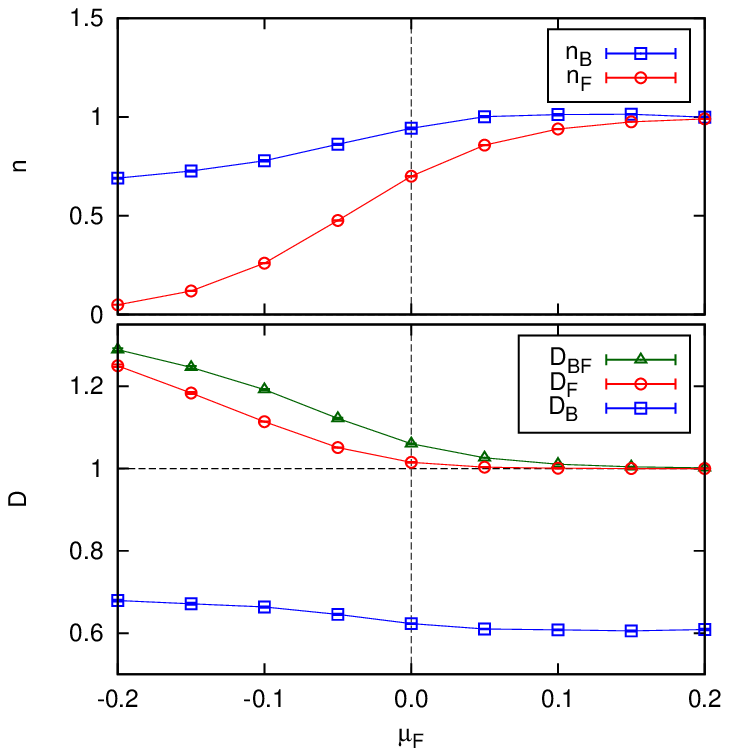}
\caption{\label{fig4}
The particle profiles as a function of $\mu_F$
for  attractive boson-fermion interaction $U_{BF}=-0.05$.
Other parameters are the same with Fig.~\ref{fig3}.
}
\end{center}
%\end{figure}

%\begin{figure}[t]
\begin{center}
\includegraphics[scale=1]{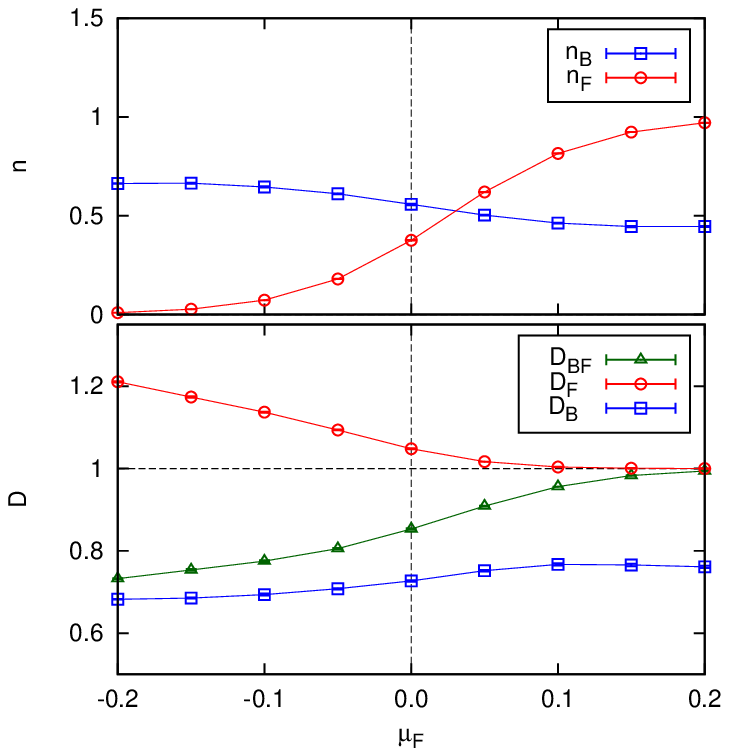}
\caption{\label{fig5}
The particle profiles 
as a function of $\mu_F$ for repulsive boson-fermion interaction $U_{BF}=+0.05$.
Other parameters are the same with Fig.~\ref{fig3}.}
\end{center}
\end{figure}

In Fig.~\ref{fig3}, we show the particle number densities ($n_B$ and $n_F$) and the 
pair occupancies ($D_{BF}$, $D_F$, and $D_B$) as a function of $U_{BF}$.   
 The chemical potentials are fixed at $\mu_B = \mu_F = 0$ as in the previous 
figures, while the temperature is chosen to be $T=0.05$ (upper boundary
  in Fig.~\ref{fig2}). 
As the boson-fermion attraction becomes stronger,
  both $n_B$ and $n_F$ increase, which is consistent with the 
  behavior expected from the effective chemical potentials 
 as we have already mentioned. Also, the pair occupancy of different (same) species
  tends to be enhanced (suppressed) as $U_{BF}$ increases, which is 
  naturally expected from the contact nature of the boson-fermion interaction.
 Note that    $D_B$ is always below unity, since we have taken repulsive
  boson-boson interaction.

In Figs.~\ref{fig4} and \ref{fig5}, we show the particle profiles as a function of the 
fermion chemical potential $\mu_F$ for
 characteristic boson-fermion interactions; an attractive case,
  $U_{BF}=-0.05$, and a repulsive case, $U_{BF}=+0.05$.
 Other parameters are the same as those adopted in  Fig.~\ref{fig3}.
 By varying $\mu_F$ in the interval [$-0.2,+0.2$],
  the fermion number density $n_F$ 
  is changed  from zero (i.e. complete vacancy) to unity (i.e. full occupancy).
The boson number density $n_B$ is an increasing (decreasing) function of
 $\mu_F$ in the attractive
(repulsive) case, which is consistent with the behavior expected from the
 effective chemical potential  $\mu_B^{\rm eff}$ in
 Eq.~(\ref{eq:muBeff}).
 Also, the boson-fermion pair occupancy $D_{BF}$ is reduced (enhanced) from unity 
in the repulsive (attractive) case as naturally expected from the contact nature of the 
 boson-fermion interaction. The fermion number density approaches to unity,
 when $\mu_F^{\rm eff}$ in Eq.~(\ref{eq:muFeff})
becomes large  either by large attractive boson-fermion interaction or by
 large  fermion chemical potential.
 If fermions fully occupy all the lattice sites, 
  both $D_{BF}$ and $D_F$ become unity, since the numerators
  of Eqs.~(\ref{eq:DBF}) and (\ref{eq:DFF}) factorize.

Although the bare fermion-fermion interaction is switched off in our present simulations
 ($U_{FF}=0$),
 there arises induced fermion-fermion attraction due to the coupling of the fermions to the 
  boson density fluctuation with the magnitude of $O(U_{BF}^2/U_{BB}$) \cite{Pethick:2008}. Because of this,
 the fermion pair occupancy $D_F$ is enhanced irrespective of the sign
 of $U_{BF}$ as seen in Figs.~\ref{fig4} and \ref{fig5}.
 Also, the boson pair occupancy $D_B$ 
 changes slightly because the boson number density $n_B$ changes. 
The boson pair occupancy $D_B$ approaches to unity as the boson becomes dilute.

\section{Summary}

In summary, we have performed a quantum Monte Carlo simulation of Bose-Fermi mixture at finite temperature  on the three-dimensional optical 
lattice. The sign problem from the boson hopping term is avoided by taking only the 
 lowest Matsubara mode, which is valid near the critical point or at high temperature.
 We have analyzed the off-diagonal long-range order of the boson, the particle number densities,
and pair occupancies for attractive and repulsive boson-fermion interaction.
We found that the region of the Bose-Einstein condensation is extended as the boson-fermion attraction  becomes strong.  The particle number profiles obtained from the simulations
 can be understood in terms of the effective chemical potentials 
 and induced fermion-fermion interaction driven by the  boson-fermion interaction.
 
 The present exploratory study is a starting point of more expanded simulations
  with the ingredients including (i) the bare fermion-fermion interaction, (ii) finite size scaling analysis with several different
   lattice volumes, (iii) continuum limit in the temporal direction,
   (iv) taking into account the non-zero Matsubara mode for bosons,
  and  (v) mapping of the phase diagram to the fixed number of bosons and fermions.
  From the physics point of view, there are various interesting future problems
   to be examined with our approach, e.g. the superfluidity of fermion pairs induced by the 
  density fluctuation of the background bosons with the boson-fermion coupling, and
  the properties of the boson-fermion mixture in the strongly coupled regime.

\section*{Acknowledgments}
The authors thank Takashi Abe and Ryoichi Seki for useful discussions.
A.~Y.~is supported by the Special Postdoctoral Research Program of RIKEN.
T.~H.~is partially supported by JSPS Grants-in-Aid  No.22340052. 
The lattice QCD simulations were carried out on NEC SX-8R in Osaka University.


\begin{thebibliography}{99}

\bibitem{Pethick:2008}
C. J. Pethick and H. Smoth, {\it Bose-Einstein Condensation in Dilute Gasses,} (Cambridge Univ. Press, London, 2008).

\bibitem{Maeda:2009ev} 
  K.~Maeda, G.~Baym and T.~Hatsuda,
  %``Simulating dense QCD matter with ultracold atomic boson-fermion mixtures,''
  Phys.\ Rev.\ Lett.\  {\bf 103}, 085301 (2009).
%  [arXiv:0904.4372 [cond-mat.quant-gas]].
  
\bibitem{Gunter:2006} 
K. Gunter, T. Stoferle, H. Moritz, M. Kohl, and T. Esslinger,
%``Bose-Fermi Mixtures in a Three-dimensional Optical Lattice,''
Phys. Rev. Lett. {\bf 96}, 180402 (2006).

\bibitem{Ospelkaus:2006}
C. Ospelkaus, S. Ospelkaus, L. Humbert, P. Ernst, K. Sengstock, and K. Bongs,
%``Ultracold heteronuclear molecules in a 3D optical lattice''
Phys. Rev. Lett. {\bf 97}, 120402 (2006).

\bibitem{Best:2009zz} 
  T.~Best, S.~Will, U.~Schneider, L.~Hackermuller, D.~-S.~Luhmann, D.~van Oosten, and I.~Bloch,
  %``Role of Interactions in Rb-87 - K-40 Bose-Fermi Mixtures in a D-3 Optical Lattice,''
  Phys.\ Rev.\ Lett.\  {\bf 102}, 030408 (2009).
  %%CITATION = ARXIV:0807.4504;%%

\bibitem{Sugawa:2011}
S. Sugawa, K. Inaba, S. Taie, R. Yamazaki, M. Yamashita, and Y. Takahashi,
%``Interaction and filling induced quantum phases of dual Mott insulators of bosons and fermions''
Nature Physics {\bf 7}, 642 (2011).

\bibitem{Takeuchi:2006}
Y. Takeuchi and H. Mori,
%``Mixing-demixing transition in one-dimensional boson-fermion mixtures''
Phys. Rev. A {\bf 72}, 063617 (2005);
%``Mixing-Demixing transition in 1D boson-fermion mixture at low fermion densities''
J. Phys. Soc. Jpn. {\bf 74}, 3391 (2005); 
%``Mott transition in one-dimensional boson-fermion mixtures''
Int. J. of Mod. Phys. B {\bf 20}, 617 (2006).

\bibitem{Pollet:2006}
L. Pollet, M. Troyer, K. van Houcke, and S. M. A. Rombouts,
%``Phase diagram of Bose-Fermi mixtures in one-dimensional optical lattices,''
Phys. Rev. Lett. {\bf 96}, 190402 (2006).

\bibitem{Sengupta:2007}
P. Sengupta and L. P. Pryadko,
%``Quantum degenerate Bose-Fermi mixtures on one-dimensional optical lattices''
Phys. Rev. B {\bf 75}, 132507 (2007).

\bibitem{Hebert:2007}
F. Hebert, F. Haudin, L. Pollet, and G. G. Batrouni,
%``Mott insulators and correlated superfluids in ultracold Bose-Fermi mixtures''
Phys. Rev. A {\bf 76}, 043619 (2007)

\bibitem{Pollet:2008}
L. Pollet, C. Kollath, U. Schollwock, and M. Troyer,
%``Mixture of bosonic and spin-polarized fermionic atoms in an optical lattice''
Phys. Rev. A {\bf 77}, 023608 (2008)

\bibitem{Varney:2008}
C. N. Varney, V. G. Rousseau, and R. T. Scalettar,
%``Quantum Monte Carlo study of the visibility of one-dimensional Bose-Fermi mixtures''
Phys. Rev. A {\bf 77}, 041608 (2008).

\bibitem{Masaki:2008}
A. Masaki, S. Tsukada, and H. Mori,
%``Mott Plateau in Trapped Bose-Fermi Mixture''
J. Phys. Soc. Jpn. {\bf 77}, 054301 (2008).

\bibitem{Pollet:2012}
L. Pollet, 
%"Recent developments in Quantum Monte-Carlo simulations with applications for cold gases"
Rep. Prog. Phys. {\bf 75}, 094501 (2012).

\bibitem{Hirsch:1983wk} 
  J.~E.~Hirsch, R.~L.~Sugar, D.~J.~Scalapino, and R.~Blankenbecler,
  %``Monte Carlo Simulations Of One-dimensional Fermion Systems,''
  Phys.\ Rev.\ B {\bf 26}, 5033 (1982).
  %%CITATION = PHRVA,B26,5033;%%

\bibitem{Prokofev:1998} 
N. V. Prokofev, B. V. Svistunov, and I. S. Tupitsyn,
JETP {\bf 87}, 310 (1998).

\bibitem{Sandvik:1999}
A. W. Sandvik,
%``Stochastic series expansion method with operator-loop update''
Phys. Rev. B {\bf 59}, 14157 (1999).

\bibitem{Rothe:2005nw} 
  For a textbook, H.~J.~Rothe,
  {\it Lattice gauge theories: An Introduction,}
  World Sci.\ Lect.\ Notes Phys.\  {\bf 74}, 1 (2005).
  %%CITATION = 00327,74,1;%%

\bibitem{Yamamoto:2012bj} 
  A.~Yamamoto,
  %``Lattice simulation of ultracold atomic Bose-Fermi mixtures,''
  Proc.~Sci. LATTICE2012, 049 (2012), in press [arXiv:1207.0376 [hep-lat]].
  %%CITATION = ARXIV:1207.0376;%%

\bibitem{Hasenfratz:1983ba} 
  P.~Hasenfratz and F.~Karsch,
  %``Chemical Potential on the Lattice,''
  Phys.\ Lett.\ B {\bf 125}, 308 (1983).
  %%CITATION = PHLTA,B125,308;%%

\bibitem{Chen:2003vy} 
  J.~-W.~Chen and D.~B.~Kaplan,
  %``A Lattice theory for low-energy fermions at finite chemical potential,''
  Phys.\ Rev.\ Lett.\  {\bf 92}, 257002 (2004).


\bibitem{Ferrenberg:1988yz} 
  A.~M.~Ferrenberg and R.~H.~Swendsen,
  %``New Monte Carlo Technique for Studying Phase Transitions,''
  Phys.\ Rev.\ Lett.\  {\bf 61}, 2635 (1988).
  %%CITATION = PRLTA,61,2635;%%
  %%CITATION = HEP-LAT/0308016;%%

\bibitem{Scalapino:1981}
D. J. Scalapino and R. L. Sugar, 
%"Monte Carlo calculations of coupled boson-fermion systems. II"
 Phys.\ Rev.\ B {\bf 24}, 4295 (1981).

\bibitem{Duane:1987de} 
  S.~Duane, A.~D.~Kennedy, B.~J.~Pendleton and D.~Roweth,
  %``Hybrid Monte Carlo,''
  Phys.\ Lett.\ B {\bf 195}, 216 (1987).
  %%CITATION = PHLTA,B195,216;%%

\end{thebibliography}
\end{document}